\documentclass{IEEEcsmag}

\usepackage[colorlinks,urlcolor=blue,linkcolor=blue,citecolor=blue]{hyperref}

\usepackage{upmath}
\usepackage{listings}
\usepackage{xcolor}

\jvol{XX}
\jnum{XX}
\paper{8}
\jmonth{March/April}
\jname{CiSE (accepted)}
\pubyear{2021}

\begin{document}

\sptitle{CiSE Special Issue (accepted)}
\editor{Publisher: IEEE}

\title{Jupyter-enabled astrophysical analysis using data-proximate computing platforms}

\author{S. Juneau}
\affil{NSF's NOIRLab}

\author{K. Olsen}
\affil{NSF's NOIRLab}

\author{R. Nikutta}
\affil{NSF's NOIRLab}

\author{A. Jacques}
\affil{NSF's NOIRLab}

\author{S. Bailey}
\affil{{}Lawrence Berkeley National Lab}

\markboth{Department Head}{Paper title}

\begin{abstract}
The advent of increasingly large and complex datasets has fundamentally altered the way that scientists conduct astronomy research. The need to work closely to the data has motivated the creation of online science platforms, which include a suite of software tools and services, therefore going beyond data storage and data access. We present two example applications of Jupyter as part of astrophysical science platforms for professional researchers and students. First, the Astro Data Lab is developed and operated by NOIRLab with a mission to serve the astronomy community with now over 1,400 registered users. Second, the Dark Energy Spectroscopic Instrument (DESI) science platform serves its geographically-distributed team comprising about 900 collaborators from over 90 institutions. We describe the main uses of Jupyter and the interfaces that needed to be created to embed it within science platform ecosystems. We use these examples to illustrate the broader concept of empowering researchers and providing them with access to not only large datasets but also cutting-edge software, tools, and data services without requiring any local installation, which can be relevant for a wide range of disciplines. Future advances may involve science platform networks, and tools for simultaneously developing Jupyter notebooks to facilitate collaborations.
\end{abstract}

\maketitle

\chapterinitial{In astronomy}, and other disciplines, the need for online science platforms is driven by both rapidly increasing data volume and by the complexity of datasets, which require highly specialized and diverse software libraries to be co-located with the data. In terms of data rates, 1 TB/night is typical of current data-intensive facilities, while the incipient Rubin Legacy Survey of Space and Time (Rubin LSST) will collect 20 TB of raw data and send a stream of 10 million alerts\footnote{Alerts are sent when a change in brightness is detected by the software comparing the new observation of a given object to previous measurements so that researchers can investigate the cause (e.g., new supernovae, variable stars).} of brightness changes each night. The size of teams working collaboratively on data analysis is also growing, routinely comprising tens to several hundreds of researchers and students. Clear advantages arise from being able to connect Jupyter Notebooks or JupyterLab to large datasets to perform analysis and data visualization efficiently both to support collaborative work and to empower a wide array of researchers including from institutions that may lack local data storage or compute resources. 

Historically, observational research was primarily conducted by individuals and small teams able to use a telescope or satellite to make observations. They would process and analyze the data using software packages on their local computers, before publishing results in astronomical journals. Larger teams and collaborations started changing the landscape by adding a need to share data among several people. Another big change came with publicly accessible data archives, which can be used to address new science questions without new observations. Researchers can download the needed data, and process and/or analyze them. For instance, statistics from the Hubble Space Telescope show that discoveries from archival data outpace those made for the original intended purpose of the observations\footnote{\url{https://archive.stsci.edu/hst/bibliography/pubstat.html}}.
Even more recently, another major change took place as some data sets are becoming prohibitively large or complex such that researchers no longer have the means to efficiently download the data onto their local computers. Instead, one needs to bring the computing workflows to the data.

Below, we showcase two astronomy projects as ongoing successful example applications. First, the Astro Data Lab (\cite{Fitzpatrick+2014,Nikutta+2020}) at NSF’s National Optical-Infrared Astronomy Research Laboratory (NOIRLab) is an online astronomy science platform currently serving large public astronomical datasets including databases with tables ranging from 500,000 to 65 billion rows. Data-proximate analysis and visualization is key to use databases containing over 100 billion rows; transferring even 150,000,000 rows with a small subset of the columns would take $\sim$2 hours over a 14 Mb/s connection, compared to $\sim$10 seconds over NOIRLab's internal 10 Gb/s connection.  It is even more relevant when considering that the databases are supplemented with Petabytes of astronomical images. While this calculus will change with availability of 5G networks, internal network upgrade opportunities will favor the code-to-data approach for the foreseeable future. Since opening its doors in 2017, the Data Lab now has over 1,500 registered users from different countries, ranging from students to faculty/researchers, and educators. 

Second, we describe how the Dark Energy Spectroscopic Instrument (DESI; \cite{Desi+2016a,Desi+2016b}) team employs Jupyter as the primary way to enable their geographically distributed collaboration (nearly 900 researchers from more than 90 institutions) to access petabytes of data with pre-installed software libraries. Jupyter Notebooks are also used to train new team members, and as a quick way to perform integration testing of software releases. For each science platform, we give an overview before presenting use cases for research, training, and education. We will then consider common elements that may apply to science platforms in general, and possible improvements from enhanced Jupyter capabilities.

\section{ASTRO DATA LAB}

\subsection{Overview and implementation}
The goal of the Astro Data Lab is to enable efficient archival use of massive astronomical survey data, particularly those produced on the facilities of NSF's National Optical-Infrared Astronomy Research Laboratory (NOIRLab).  As shown in {\bf Figure 1}, the data volume from NOIRLab facilities has grown explosively over the last decade, with much of that growth owed to wide-angle images taken with the Dark Energy Camera\footnote{\url{https://www.darkenergysurvey.org/the-des-project/instrument/}}.  These surveys of the sky result in large maps with a multitude of stars, as well as galaxies that are billions of light-years from Earth. These maps are typically processed through automated measurement pipelines, yielding tables of billions of astronomical objects. The value of these processed data is enormous, not just for the teams taking the data for their own purposes; much of the value lies in the potential for the broad astronomical community, and the public, to make unanticipated discoveries.

Astro Data Lab enables several kinds of archival science cases, including 1) discovery of unusual, rare, or otherwise interesting objects selected through queries to databases; 2) classification of objects in large samples, potentially including image cutouts and aided by machine learning techniques; and 3) discovery in the time domain, where time series data of selected objects are used to create new understanding or reveal new phenomena.  All of these science cases are supported by services provided by Astro Data Lab, with Jupyter Notebooks as the key technology that brings their power to users.  For example, users can query databases directly within a Notebook (via a {\it QueryClient}), and save their output to storage associated with their Data Lab accounts, which is currently one Terabyte per user (via a {\it StoreClient}). {\bf Figure~2} illustrates how these pieces connect together to make Jupyter an integral part of the Astro Data Lab science platform, and we include examples of the code syntax to use the Clients within a Notebook in the Appendix.

\begin{figure*}
\centerline{\includegraphics[width=29pc]{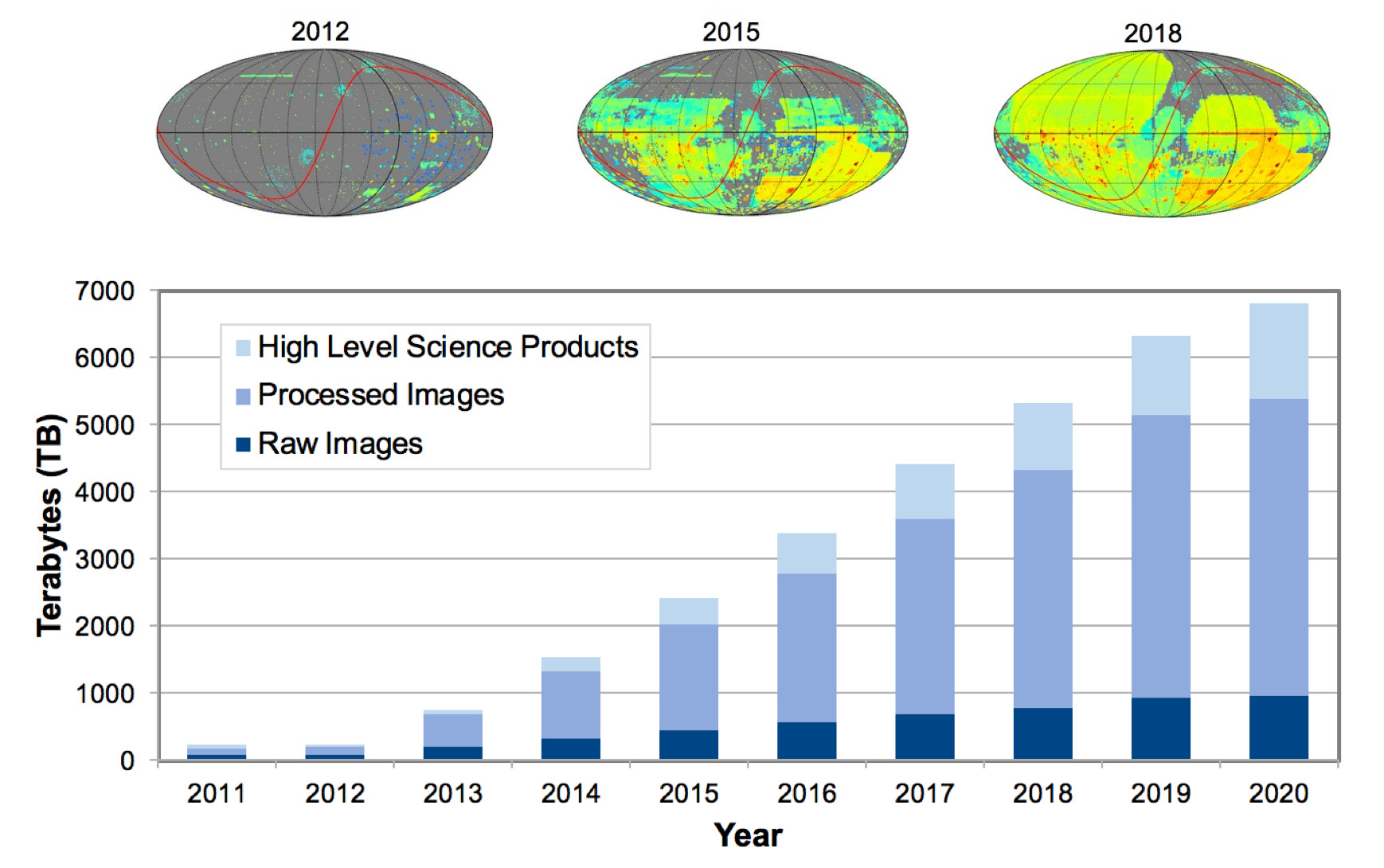}}
\caption{Evolution of the sky coverage from NOIRLab's observatories with telescope diameter of 2-4 meters during the years 2012, 2015 and 2018. The color scheme indicates the exposure time on a logarithmic scale, mostly ranging from 900~seconds (15 minutes; green) up to nearly 1 million seconds (277~hours; dark red). The rapid increase of coverage starting around 2013 coincides with astronomical surveys conducted with wide-field cameras such as the Dark Energy Camera (DECam). These full-sky maps were generated with an equal-area {\it Mollweide} projection, and the Milky Way Galactic Plane is shown with a solid red line. The bar chart illustrates the growth in data volume stored at NOIRLab's Astro Data Archive, split between unprocessed (raw) images, processed images, and high-level data products such as image mosaics.}
\end{figure*}

The Astro Data Lab science platform runs on premises using commodity hardware (1U and 2U general-purpose servers with typically two Xeon CPUs and ~20 cores each). The various data services and the Jupyter notebook server are currently deployed as virtual machines, with the available resources shared by all users. To facilitate resource allocation, and permit users to maintain their own software stack, we are currently working to implement containerization based on Docker and Kubernetes. Specialized hardware consisting of three machines (production, hot standby, staging) is deployed for the catalog databases, and is optimized for CPU/RAM throughout and backed by fast SSD storage.

\begin{figure*}
\centerline{\includegraphics[width=30pc]{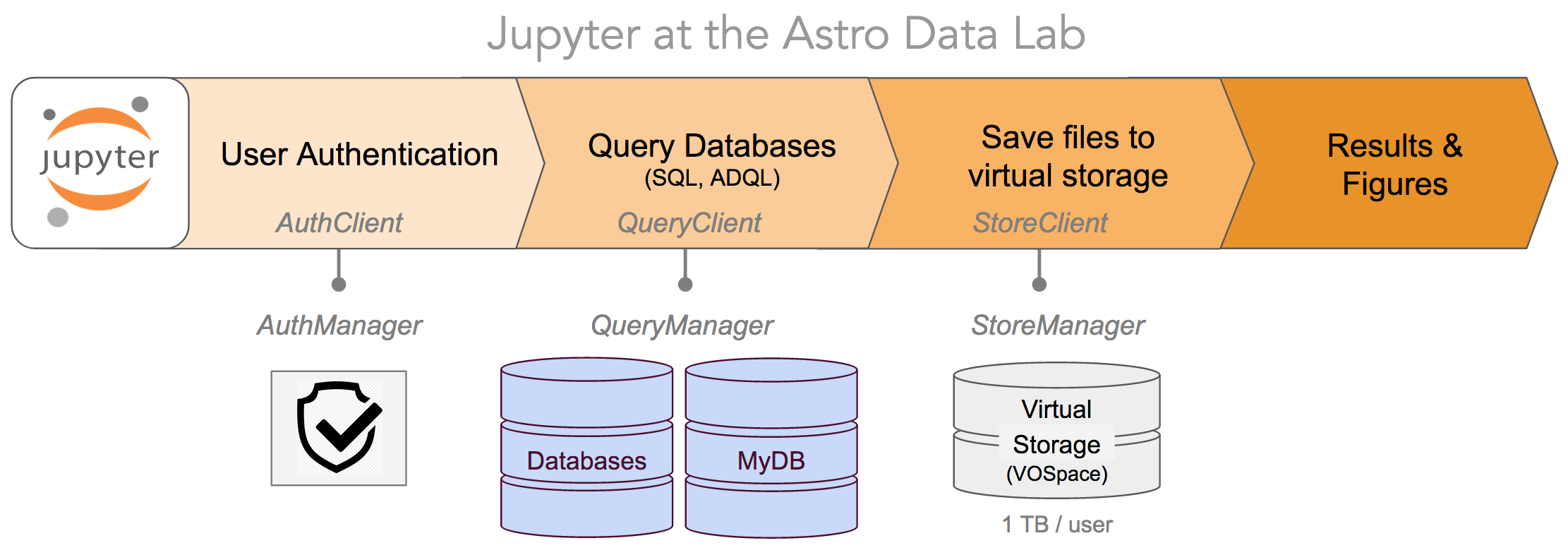}}
\caption{Illustration of some of the key Clients and Managers that are part of the Data Lab platform, and which can be used with Jupyter to connect the users to the data, and enable them to conduct their analysis using Notebooks. User authentication is only required for users to access their MyDB or virtual storage space. The queryClient allows users to send a database query, and retrieve the results directly within a Notebook. The overall goal is for users to be able to conduct their entire analysis up to generating publication-quality figures on the Jupyter server.}
\end{figure*}\label{fig:jupdl}

\subsection{Jupyter for Research}

The majority of Astro Data Lab users are professional astronomers and graduate students. They tend to have topical expertise in their subject of research but varied levels of coding experience and knowledge. To remove the need for local installation, the Jupyter Notebook server includes pre-installed software packages and Python libraries (e.g., numpy, matplotlib, pandas), and is populated with example Jupyter notebooks directly within the user accounts. The default installation includes astronomy-specific packages such as Astropy\footnote{\url{https://www.astropy.org/}} (\cite{Astropy+2018}) and the datalab clients\footnote{\url{https://github.com/noaodatalab/datalab}}.

The Astro Data Lab Jupyter notebook server provides a convenient way to quickly query database tables, visualize their content, and perform detailed analyses. As an example scientific use case, we consider the Survey of the Magellanic Stellar History (SMASH\footnote{\url{https://datalab.noirlab.edu/smash/smash.php}} \cite{Nidever+2017}), which mapped a wide area in and around the periphery of the Magellanic Clouds, two of the nearest companion galaxies to the Milky Way. The SMASH database table contains millions of stars forming the Magellanic Cloud galaxies as well as millions of distant, background galaxies, totaling nearly 360 million unique objects. Discoveries from SMASH include a ripple in the disk plane of the Large Magellanic Cloud (LMC; \cite{Choi+2018}) $-$ likely the result of an interaction with its neighbor galaxy the Small Magellanic Cloud $-$ and a previously unknown faint dwarf galaxy companion to the Milky Way, named Hydra II (\cite{Martin+2015}). The study of such faint companion galaxies may hold clues to the nature of the unknown ``dark matter''.

The Hydra II dwarf galaxy finding is featured in a science example Jupyter Notebook provided to users, giving them a demonstration of a discovery from beginning to end, and a starting point for their own explorations.  It shows the chain of steps for research from how the database query is constructed and issued to the server, how to display the measurements that are returned, and how to process them to highlight the newly discovered galaxy. 
Researchers and students can also browse other example notebooks from within their Astro Data Lab account, or on the ``03\_ScienceExamples'' GitHub repository\footnote{\url{https://github.com/noaodatalab/notebooks-latest/tree/master/03\_ScienceExamples}}.

Researchers who conduct their analysis on the Astro Data Lab platform are asked to include a formal acknowledgement in all resulting journal publications, and to use the {\it facilities} command ``Astro Data Lab'' when available, such as for the American Astronomical Society journals. They may further propose to add their Jupyter Notebook to our contributed Notebook collection\footnote{\url{https://github.com/noaodatalab/notebooks-latest/tree/master/05\_Contrib}}. User-contributed notebooks that show applications not yet covered in our collection are especially welcome. The ``05\_Contrib'' folder is divided into astronomy topics, and includes step-by-step instructions as well as an Astro Data Lab Notebook template for guidance. The team then reviews the proposed notebook, and may iterate with the contributor until the Notebook is approved to be added to the notebook collection, which will consist in merging the pull-request (PR) on GitHub. The vetting process requires manual interventions and topical knowledge. Once the GitHub PR is merged, the distribution and testing of the new contributed notebook is automated as described below.

\subsection{Jupyter for Training}

Training is crucial at all career levels. This is partly due to the important changes on the software, tools, and methods being developed and utilized. For example, several experienced researchers are newly learning the Python programming language. More fundamentally, this is also due to the increasing prominence of data-driven modes of conducting science. Training presents new challenges as we need to convey and teach new, data-intensive methods, but it also presents new opportunities. 
Data Lab in combination with Jupyter helps make research fairer and more equitable. Improving access to quality data is an important part of building an inclusive community by lowering the barrier of entry, thus leveling the playing field for all (\cite{Norman2018}).

The Astro Data Lab recognizes the importance of providing researchers, students, and professors with adequate technical training and support in data science to empower them to take advantage of the increasing volume and complexity of datasets. In terms of documentation and tutorials, the team produced a User Manual\footnote{\url{https://datalab.noirlab.edu/docs/manual/index.html}} using Sphinx Documentation\footnote{\url{https://www.sphinx-doc.org}} as well as numerous example Jupyter Notebooks, which are available as a reference at any time. In addition, team members give live presentations and webinars, conduct online and/or in-person demos as well as hands-on tutorials, during which participants can interactively follow along. Most of the demos and tutorials involve going through Jupyter Notebooks, which range from entry-level to advanced science examples to technical "How-To" notebooks, and which are described in a README file. The procedure to maintain, and automatically update the collection of Notebooks is described below.

\subsection{Jupyter for Education}

Jupyter Notebooks facilitate education about topics in astronomy and/or data science, including computer coding, data manipulation and analysis. We currently support the Python language, and have two primary sets of notebooks under the Education and Public Outreach (EPO) umbrella. 

The Teen Astronomy Cafe Program\footnote{\url{http://www.teenastronomycafe.org/}} consists of hands-on computer activities and discussions led by volunteer scientists in addition to the program director. The target audience is 12 to 18 year-old students. Jupyter Notebooks are produced ahead of time, and may include interactive widgets (sliders, and buttons) and/or instructions to teach students to modify code cells. Each activity is designed to cover a different astronomy topic for which data science techniques can be employed. As of 2020, these EPO Notebooks can be run either on Google Colab or directly on the Astro Data Lab platform.

At the level of undergraduate and graduate students, the Astro Data Lab is adding a collection of notebooks adapted from the La Serena School for Data Science\footnote{\url{http://www.aura-o.aura-astronomy.org/winter_school/}}. These notebooks were originally written by the guest lecturers, and cover more advanced topics involving, e.g., statistical tools applied to large datasets, high performance computing, image analysis, and machine-learning techniques. The goal is for students to learn hands-on skills that they can then apply to their own research.

Lastly, Data Lab notebooks have been used in the classroom upon requests from professors and teachers, including semester-long graduate-level astronomy courses. In some instances, we create temporary {\it demo} accounts for students. These in-house demo accounts are generated via a script, and allow the users to save their work in progress during the duration of the fixed-term account. After they expire, they are wiped clean, and deactivated using an automated script.

\subsection{Notebook Deployment and Testing}

The Data Lab team develops a suite of notebooks for all users. These range from notebooks introducing Python, SQL, and Data Lab, over technical How-Tos, to complete science cases that reproduce 
interesting results from the astronomical literature. Astronomy and data
science education notebooks complete the picture. Data Lab also
assists users who wish to contribute a notebook to the collection by providing them with a template and step-by-step instructions.

Notebooks are developed either on the Data Lab notebook server\footnote{\url{https://datalab.noirlab.edu/devbooks/}}, or locally (with the
\texttt{datalab} package\footnote{\url{https://github.com/noaodatalab/datalab/}} installed). All
version control occurs through GitHub\footnote{ \url{https://github.com/noaodatalab/notebooks-latest/}}.
Pull requests are reviewed by the Data Lab team. When merged to the master
branch, a webhook is triggered which sends a beacon to the Data Lab
servers. If the beacon is successfully verified, the server pulls the
latest version from GitHub and re-populates a specific directory with
the updated notebooks. This \texttt{notebooks-latest/} directory is
read-only mounted in all users' notebook spaces, and it can be browsed
both using a browser and a terminal. Users can also pull a whole new
read \& write copy of it using a custom shell command
(\texttt{getlatest <targetdir>}).

The notebook suite is now around 40 notebooks strong and growing. We
must ensure that all notebooks continue to execute correctly at all
times, despite the backend datasets and middleware services being
updated often. To facilitate automated pass/fail testing we employ
\texttt{nbconvert} wrapped in a custom Python package. Testing can be
performed through a dedicated notebook (part of the notebook suite),
or in command line mode. The testing supports globbing for
\texttt{.ipynb} files, exclusion patterns, dynamic notebook kernel
switching, and summary reporting. Test failures do not interrupt the
testing suite, and the tracelog of all tested notebooks is available
for inspection. Currently the test suite completes in 35 minutes on
the Data Lab notebook server.

\section{DESI}

\subsection{Overview and implementation}

The DESI instrument (\cite{Desi+2016b}) will be used to survey one third of the sky (14,000 square degrees), and obtain spectra for 35 million galaxies and quasars in order to construct the largest three-dimensional map of the universe to date \cite{Desi+2016a}. 
By comparing this map to predictions from numerical models, the DESI team will constrain the nature of dark energy, the dominant yet mysterious force driving the expansion of the universe. In addition to its main cosmology goals, which rely on spectroscopy of very distant galaxies and quasars, DESI will obtain spectra for 10 million stars from the Milky Way, and an additional 10 million brighter, less distant galaxies. These added-value data will allow astronomers to answer a slew of astrophysical questions about stars, galaxies, and their evolution.

Currently installed on the four-meter Mayall telescope at Kitt Peak National Observatory, Arizona, DESI can obtain spectra of up to 5000 targets in a single observation, by allocating a single optical fiber per target. 
The DESI team first needs to measure the sky positions (coordinates) of all the galaxies, quasars, and stars to observe with DESI, select which ones can be observed simultaneously with 5000 fibers, and compute how to align each {remotely-controlled} fiber with its target ({\it fiber assignment}). The targets are selected from the pre-imaging Legacy Surveys\footnote{\url{https://www.legacysurvey.org/}}, which are now complete, and consist of optical and infrared images which were used to measure the position and fluxes of over 1.6 billion unique objects.

DESI generates $\cal{O}$(1 PB) of data per year, including raw telescope data, processed results, and computer simulations needed for final science analyses. This amount of data together with the sophisticated and specialized software were motivators to use a Jupyter server running close to the data at the National Energy Research Scientific Computing Center (NERSC), where team members can sign in and launch an instance of JupyterLab with pre-installed software.  The JupyterHub servers are maintained by NERSC as a service to all users, not specific to DESI, but they are significantly used by DESI collaborators using DESI-maintained custom kernels with DESI-specific software.

NERSC deploys its JupyterHub service using a container-as-a-service platform based on Kubernetes, and uses wrapspawner\footnote{\url{https://github.com/jupyterhub/wrapspawner}} and batchspawner\footnote{\url{https://github.com/jupyterhub/batchspawner}} to enable JupyterLab sessions to spawn on shared or exclusive (batch) nodes on CPU or GPU resources.  Most DESI collaborators using Jupyter access it via a node shared with other NERSC users for non-CPU-intensive exploratory analysis and visualization.  More intensive Jupyter-based computing is available through distributed analytics platforms like Spark or Dask, or MPI-enabled applications leveraging IPyParallel, managed from notebooks through script-based adaptations to high-performance computing environments.

Below, we present example use cases of Jupyter for the DESI collaboration of nearly 900 members including research staff, postdoctoral researchers, graduate students and engineers. These include examples for the purpose of research, training, education, and software testing. The DESI software is maintained on GitHub (\texttt{desihub}\footnote{\url{https://github.com/desihub}}), and includes 59 repositories with 107 members.

\subsection{Jupyter for Research}

In the case of DESI, research includes survey planning, and data processing, exploration, and analysis. The research applications themselves are fairly broad as they include both the primary cosmology goals centered around dark energy as well as a slew of possible other projects on galaxies, stars, black holes, etc.  Therefore, having a suite of tools and analysis workflows will maximize the outcome from the DESI observations.  Jupyter enables interactive exploration and analysis of petabytes of data by remote collaborators without requiring them to download the data locally.  It also provides kernels with pre-installed software releases for DESI-specific data analysis.

Jupyter is frequently used by DESI collaborators for initial data exploration and algorithm development ({\bf Figure~3}), before deploying those algorithms as Python packages to be used in non-Jupyter batch jobs using hundreds to thousands of nodes.  Jupyter notebooks are also used for collating the results of those batch jobs and visualizing the final results.

A specific example of data exploration is the interactive spectral visualization tool \texttt{prospect}\footnote{\url{https://github.com/desihub/prospect}}, which is developed on GitHub, and used by the DESI team at NERSC to visually inspect DESI spectra to debug data processing problems. It is based on Bokeh, and works in the JupyterLab environment. {\bf Figure~4} gives an overview of a version of \texttt{prospect} that is being implemented at the Data Lab, showing an example application with a Sloan Digital Sky Survey (SDSS) spectrum of a galaxy. 

\begin{figure*}
\centerline{\includegraphics[width=31pc]{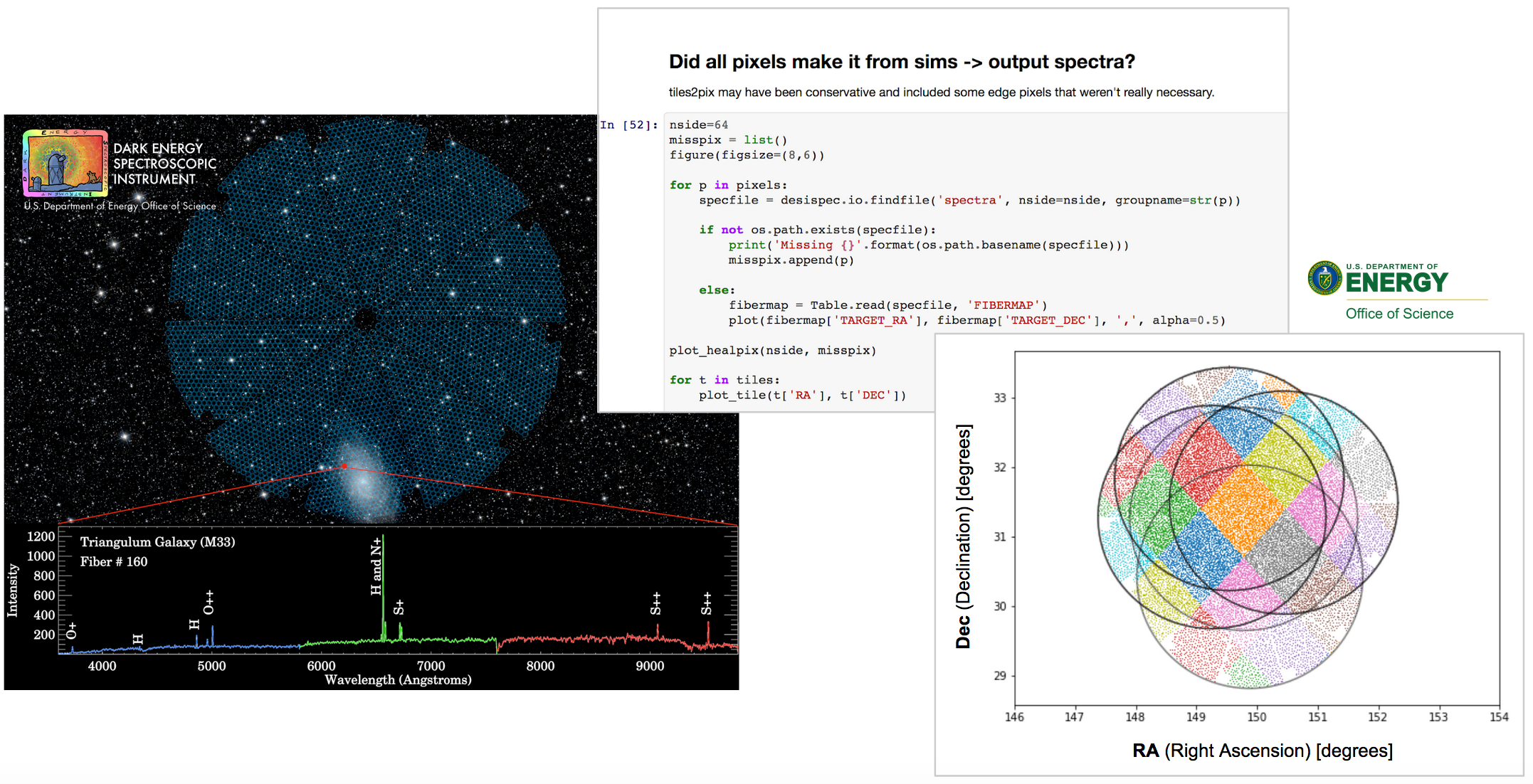}}
\caption{ {\it (Left)} Overlay of the DESI focal plane on the sky with 5000 fiber patrol regions. The example test spectrum at the bottom was obtained from a fiber positioned on a part of the Triangulum Galaxy (red dot) as part of early commissioning of the instrument in October 2019, and shows spectral features as labeled (Credit: DESI Collaboration; Legacy Surveys; NASA/JPL-Caltech/UCLA); {\it (Middle)} Section of a Jupyter Notebook using the DESI software to plot the fiber location to visualize how overlapping DESI pointings can result in repeated observations, which can be used to obtain more signal. {\it (Right)} Resulting plot shown inline in the Notebook. The black circles illustrate the full size of a given DESI pointing (focal plane), while each colored dot marks the location of a fiber. The color scheme is mapped to HealPix indices, which are defined to tile the sky with nside=64 in this example. The color saturation level is keyed to the number of repeated fiber assignments (darker colors signify more overlap).}
\end{figure*}

\begin{figure*}
\centerline{\includegraphics[width=29pc]{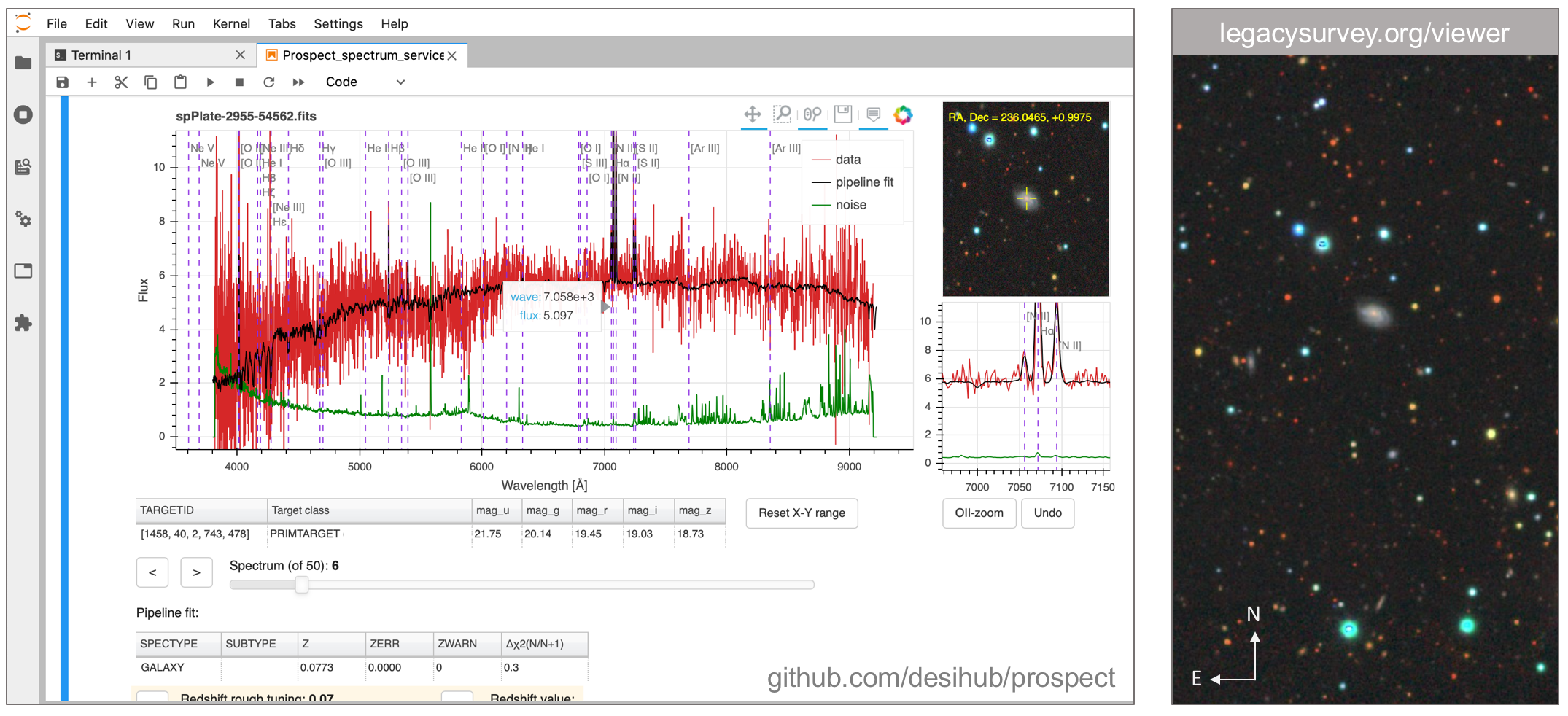}}
\caption{Interactive spectral visualization tool \texttt{prospect}, which is developed on GitHub, and used by the DESI team at NERSC. It is based on Bokeh, and works in the JupyterLab environment. Above is a version of \texttt{prospect} that is being implemented at the Data Lab, showing an example application with a SDSS spectrum of a galaxy (in red), with the best-fit model overlaid in black, and the error spectrum in green. Vertical dashed lines mark the location of expected spectral features such as emission lines from ionized gas (better seen in the bottom right {\it zoom box} inset, which zooms around the location of the cursor by capturing the mouse-over information). The target galaxy can be seen in the top right image along with its celestial coordinates in Right Ascension (RA) and Declination (Dec). The interactive tool allows users to zoom, pan, and reset the plot by using the Bokeh buttons at the top of the viewer, and displays tabulated information about the object being plotted (6th object of 50 in the example above as indicated by the slider). Clicking on the image thumbnail leads to the location of the galaxy in the interactive Sky Viewer (https://www.legacysurvey.org/viewer; screenshot on the right; with the North and East direction indicated with arrows).}
\end{figure*}

\subsection{Jupyter for Training}

Jupyter Notebooks are used to create DESI Tutorials\footnote{\url{https://github.com/desihub/tutorials}} to train new DESI collaboration members, and to teach existing members new tasks or functionalities. They are maintained and can be browsed on GitHub. However, DESI team members must run them at NERSC after authenticating, cloning the notebooks from GitHub, and launching a JupyterLab instance on the Jupyter server. The underlying DESI software is written in Python, and is pre-installed. The user has the ability to select DESI kernels to test different software versions if needed, or simply run the default kernel. Tutorials range from basic functionality to generating mock data to planning observations via fiber assignment calculations, and investigating available spectra from early observations. They are meant to train users in a modular fashion, and to enable them to re-use and combine parts from different tutorials in order to create their own workflow.

\subsection{Jupyter for Education}

One of the educational efforts using Jupyter is a program called \emph{DESI High}\footnote{\url{https://github.com/michaelJwilson/DESI-HighSchool}}, which aims to teach 14 to 18 year-old students to understand Dark Energy with actual data from the DESI project and Jupyter notebooks. The latter are made available on GitHub and can be run with Binder. The DESI High program also includes a lot of links to useful material and references such as videos, websites, tutorials. An underlying motivation of this program, which is shared with some of the initiatives mentioned previously, is to inspire and teach the next generation of (data) scientists.

\subsection{Jupyter for Software Testing}

Integration testing of DESI software releases is performed by a Jupyter Notebook on the Jupyter server at NERSC. The ``minitest.ipynb'' notebook in the GitHub \texttt{desitest}\footnote{\url{https://github.com/desihub/desitest}} repository runs an end-to-end integration test from simulating input data to processing it with the data pipeline to performing final quality assurance and analysis.
This notebook includes several steps, some of which spawn batch jobs and maintain logfiles outside of the notebook itself. Under normal activity levels on the NERSC servers, the minitest notebook runs end-to-end in about two hours and uses approximately 500 core-hours of computing resources, a scale that would not be viable for typical continuous integration services.  The output from this notebook serves as the ``reference run'' for the software release, providing specific examples of the data outputs, formats, and performance for that release. 

\section{JUPYTER IN SCIENCE PLATFORMS}

We now consider similarities and differences between the Astro Data Lab and DESI examples, as well as other astrophysical science platforms. Both the Data Lab team and the DESI team take advantage of the user-friendly format of Jupyter Notebooks to create examples and tutorials to train new users, and to illustrate key functionalities. They both maintain their tutorials and examples on GitHub for version control and contributions. In the case of Data Lab, each user account comes with the latest collection of example Notebooks which range from “Getting Started” novice level, to more technical “How To” notebooks, complete scientific use cases, as well as educational notebooks and contributed notebooks from the astronomy community. In the case of DESI, team members can follow instructions to clone the notebooks from GitHub and start using them. 

The main difference may be the target audience, in the sense that the Astro Data Lab is primarily serving publicly available data to the general astronomy community, while the DESI team has restricted access to proprietary data. Furthermore, the DESI effort is centered around the instrument's primary cosmology mission. In contrast, the Astro Data Lab aims to maximize the scientific output of the community including all topics in astronomy and/or cosmology.

\subsection{Toward a Network of Science Platforms}

Other astronomy science platforms are either active or being developed, and all rely on Project Jupyter to enable their users to perform data proximate analysis. For instance, the Space Telescope Science Institute (STScI) is utilizing Project Jupyter by 1) curating Jupyter notebooks as training materials\footnote{\url{https://github.com/spacetelescope/notebooks}} for astronomy mission data services;  2) building services for science users to discover notebooks and utilize Jupyter-based tools such as the \texttt{Jdaviz}\footnote{\url{https://jdaviz.readthedocs.io/en/latest/}} package of astronomical data analysis visualization tools, and 3) deploying science platforms in the Amazon Web Service (AWS) that provide these notebooks and compute resources to users in proximity to public mission data. STScI develops these platforms to support community engagement through workshops 
as well as through archive services, for example as an entry point for users to interact with data from the upcoming Roman Space Telescope mission.

Another example is the Canadian Astronomy Data Center (CADC\footnote{\url{http://www.cadc-ccda.hia-iha.nrc-cnrc.gc.ca/en/}}), which is currently developing a Jupyter Notebook server to be deployed on the Canadian Advanced Network for Astronomical Research (CANFAR\footnote{\url{https://www.canfar.net/en/}}) in 2021, and which will utilize authentication and storage (VOSpace\footnote{\url{https://www.ivoa.net/documents/VOSpace/}}) protocols and group permission already implemented at CADC, in addition to providing users with access to data archives, and software. 

Also in development, the Vera C. Rubin Observatory uses JupyterLab as the core technology of the Notebook Aspect of its Rubin Science Platform\footnote{\url{https://nb.lsst.io}} (RSP).  The Rubin Observatory, currently under construction, will feature a wide-field 8.4-meter telescope, a gigapixel camera, and a data management system capable of handling the hundreds of petabytes that its 10-year Legacy Survey of Space and Time (LSST) will collect.  The Notebook Aspect will be the key interface allowing the Rubin user community to access the massive dataset and run the LSST science pipelines.  Other major projects in development planning their own science platforms include the Maunakea Spectroscopic Explorer\footnote{\url{https://mse.cfht.hawaii.edu}}, an 11-meter spectroscopic survey telescope, and the U.S. Extremely Large Telescope Project, which will have a share of both the Thirty Meter Telescope\footnote{\url{https://www.tmt.org}} and the 25-m Giant Magellan Telescope\footnote{\url{https://www.gmto.org}}.

Challenges that may limit the adoption of science platforms include keeping the barrier of entry as low as possible for onboarding new users with no prior experience, the potential cost or agency limitations if platforms rely on commercial cloud solutions, and the complexity or heterogeneity of datasets and services. The latter can be exacerbated if each platform deploys different, non-standard solutions. A tighter communication across platforms and disciplines including sharing experiences, designing for inter-operability, and discussing and adopting common core technologies (e.g., Jupyter, Docker, Kubernetes) would alleviate some of these expected challenges.

While the pace of technological upheaval in software and data science is rapid, Jupyter has clearly discovered a need in the realm of astronomical research for a flexible, extensible, and easily deployable interactive software environment, and thus is well-positioned to remain a core technology.  This need is likely to expand as astronomical data volumes grow further. Indeed, it is not difficult to imagine a future in which the bulk of astronomical data analysis moves to the Cloud, in part powered by Jupyter, perhaps through a system as that envisioned by the Science Platform Network (\cite{Desai+2019}).

\section{CONCLUSION}

In this paper, we considered how Jupyter can play a central role in astrophysical science platforms by connecting large volumes of data with a local analysis host. Namely, we presented two example cases that employ a Jupyter server to provide researchers and students with access to Petabytes of data, with pre-installed software for advanced data analysis, and with tutorials. We first described NOIRLab's Astro Data Lab, which is operating based on a mission to maximize the scientific outcome of the astronomy community from publicly-available wide-field astronomical surveys. 

Beyond performing research, Astro Data Lab's efforts extend to training users and contributing to education in astronomy and data science. We then presented how the DESI team utilizes Jupyter for a range of activities such as survey preparation, software development, research, training, and education. Online science platforms share the common goal of co-locating software, tools, and tutorials with the data. This is increasingly needed as datasets grow both in volume and in complexity. 

Science platforms can be leveraged to improve on the robustness and on the reproducibility of research results by making entire analysis workflows available. For instance, Jupyter Notebooks could be added to (or cited by) peer-reviewed journal publications. In addition to posting Notebooks on an open source platform such as in a GitHub repository, one can create a Release for the repository, and register a persistent digital object identifier (DOI) with the open repository Zenodo\footnote{\url{https://zenodo.org/}}. This approach ensures the longevity of published Notebooks. Additional coordination between science platforms and professional journals may further strengthen this good practice.

Another interesting future outlook is the opportunity to facilitate collaboration between researchers. In particular, users have already often expressed their interest in being able to collaboratively write Jupyter notebooks in real-time as part of online platforms. This may be achieved with recent, ongoing developments, such as Jupyter Real-Time Collaboration (Jupyter-RTC). Such enhanced capability would further strengthen the impact of Jupyter in accelerating the progress of astrophysical discoveries.

Taking an even broader perspective, there are many parallels to draw with other scientific disciplines where similar data proximate computing platforms are developed and operated. For instance,  CyVerse\footnote{\url{https://www.cyverse.org/}} is an interdisciplinary platform heavily used for biology and bioinformatics. While the type of data differs (e.g., genome sequence rather than stars and galaxies), there are similar needs in terms of filtering, analyzing, and visualizing data. Other examples include Pangeo\footnote{\url{https://pangeo.io/}}, a platform for open, reproducible, and scalable geoscience.  Many of the challenges faced and solutions adopted by these projects are transferable to astrophysical data science, creating a wealth of opportunity for future cross-disciplinary collaboration.

\section{APPENDIX}

\noindent
We introduce how to use the Data Lab Clients within a Jupyter Notebook, which were illustrated in Figure~2. First, the authClient works by creating a token to recognize that a user is properly authenticated, which is required to access virtual storage (VOSpace) and personal database space (MyDB). This step can be skipped if a user does not need to access those in a given Notebook, or if the user has recently authenticated as the token persists in the account.

\begin{lstlisting}
from dl import authClient as ac
from getpass import getpass

token = ac.login(
          input("Enter user name:"),
          getpass("Enter password:"))
if not ac.isValidToken(token):
    raise Exception("Token is not valid. Please check your usename/password and execute this cell again.")
\end{lstlisting}

Secondly, we show how the queryClient can be used to send a SQL (and ADQL,\footnote{\url{https://www.ivoa.net/documents/ADQL/}} Astronomical Data Query Language) query and retrieve the results in a Jupyter Notebook, formatted as a pandas DataFrame in the example below, and run synchronously (the optional keyword \texttt{async\_} can be set to True to run the query asynchronously, in which case a jobId is returned, which can be used to check on the status of the query, and retrieve the results after the query has completed).

\begin{lstlisting}
from dl import queryClient as qc

query = "SELECT * FROM ls_dr8.tractor LIMIT 1000"
df = qc.query(sql=query, fmt='pandas')
\end{lstlisting}

Thirdly, we show how the storeClient can be used to store files to and retrieve from VOSpace. VOSpace is an implementation of network-attached storage, and each user account comes with a dedicated storage area. The result of a query can be written out directly to VOSpace:

\begin{lstlisting}
res = qc.query('SELECT * FROM ls_dr8.tractor LIMIT 1000', out='vos://lsdr8_sample.csv')
\end{lstlisting}

\noindent
The file can be later retrieved:

\begin{lstlisting}
from dl import storeClient as sc
from dl.helpers.utils import convert

res = sc.get('vos://lsdr8_sample.csv')
df = convert(res,'pandas')
\end{lstlisting}

\noindent
To share files with others, users can place files within the \texttt{public/} subdirectory in their VOSpace, e.g.

\begin{lstlisting}
sc.mv(fr='vos://lsd8_sample.csv', to='vos://public/lsdr8_sample.csv')
\end{lstlisting}

\noindent
Another user can then access it, and convert the content to, e.g., a pandas DataFrame:

\begin{lstlisting}
res = sc.get( 'otheruser://public/lsdr8_sample.csv')
df = convert(res,'pandas')
\end{lstlisting}

\section{ACKNOWLEDGMENT}

We thank the remainder of the Astro Data Lab team members, including A. Bolton, M. Fitzpatrick, D. Herrera, L. Huang, D. Nidever, R. Pucha, and A. Scott. In particular, we thank B. Weaver for producing Figure 4, and for his work adapting the \texttt{prospect} spectral viewer to work at the Astro Data Lab. We also thank the three reviewers for their useful comments that helped us to improve this manuscript, as well as S. Gaudet (National Research Council Canada), and G. Snyder (STSci) for sharing information about astronomical science platforms in which they are directly involved, R. Thomas (NERSC) and G. Jacoby for comments on the manuscript. Astro Data Lab is part of NSF's National Optical-Infrared Astronomy Research Laboratory (NOIRLab), which is operated by the Association of Universities for Research in Astronomy (AURA), Inc. under a cooperative agreement with the National Science Foundation. DESI is supported by the Director, Office of Science, Office of High Energy Physics of the U.S. Department of Energy under Contract No.DE–AC02–05CH1123, and by the National Energy Research Scientific Computing Center, a DOE Office of Science User Facility under the same contract; additional support for DESI is provided by the U.S. National Science Foundation, Division of Astronomical Sciences under Contract No. AST-0950945 to the NSF’s NOIRLab; the Science and Technologies Facilities Council of the United Kingdom; the Gordon and Betty Moore Foundation; the Heising-Simons Foundation; the French Alternative Energies and Atomic Energy Commission (CEA); the National Council of Science and Technology of Mexico; the Ministry of Economy of Spain, and by the DESI Member Institutions. The DESI team is honored to be permitted to conduct astronomical research on Iolkam Du’ag (Kitt Peak), a mountain with particular significance to the Tohono O’odham Nation.

\begin{IEEEbiography}{St\'ephanie Juneau,}{\,} Associate Astronomer at NSF's NOIRLab and Data Scientist at the Astro Data Lab (in Tucson, AZ). Dr. Juneau received a Ph.D. in Astronomy from the University of Arizona. Her scientific research is focused on galaxies and supermassive black holes. She is interested in applying statistical methods, advanced visualization techniques, and in improving astrophysical science platforms. 
She is a member of the DESI and Euclid collaborations, and of the American Astronomical Society (AAS). Contact her at stephanie.juneau@noirlab.edu.
\end{IEEEbiography}

\begin{IEEEbiography}{Knut Olsen,}{\,} CSDC Principal Data Scientist at NSF's NOIRLab (in Tucson, AZ).  Dr. Olsen received his Ph.D. in Astronomy from the University of Washington. He works on stellar populations in nearby galaxies. Contact him at knut.olsen@noirlab.edu.
\end{IEEEbiography}

\begin{IEEEbiography}{Robert Nikutta,}{\,} Project Scientist for Astro Data Lab at NSF's NOIRLab (in Tucson, AZ). Dr. Nikutta received his Ph.D. from the University of Kentucky. His scientific interests range from the nuclear regions of Active Galactic Nuclei to astroinformatics and big data research. Contact him at robert.nikutta@noirlab.edu
\end{IEEEbiography}

\begin{IEEEbiography}{Alice Jacques,}{\,} Data Analyst at NSF's NOIRLab and Astro Data Lab (in Tucson, AZ). Alice received her B.S. degree in Physics and Computer Science from Georgia College and State University and her M.S. degree in Physics from the University of Louisville. Contact her at alice.jacques@noirlab.edu.
\end{IEEEbiography}

\begin{IEEEbiography}{Stephen Bailey,}{\,} Senior Software Developer in the Physics Division at
Lawrence Berkeley National Lab (in Berkeley, CA).  Dr.~Bailey received his Ph.D. in Physics from Harvard University and currently leads the DESI Data Management team.  He enjoys converting raw data into useful data for the benefit of large science collaborations, and has been doing so for over 25 years.  Contact him at StephenBailey@lbl.gov.
\end{IEEEbiography}

\end{document}